\documentclass[aps,twocolumn]{revtex4}

\usepackage{graphicx,epsfig}

\def\duzomniejsze{<\kern-.7mm<}
\def\duzowieksze{>\kern-.7mm>}

\def\textbf#1{{\bf #1}}
\def\be{\begin{equation}}
\def\ee{\end{equation}}
\def\ben{\begin{eqnarray}}
\def\een{\end{eqnarray}}
\def\eea{\end{array}}
\def\bea{\begin{array}}
\newcommand{\bei}{\begin{itemize}}
\newcommand{\eei}{\end{itemize}}
\newcommand{\bee}{\begin{enumerate}}
\newcommand{\eee}{\end{enumerate}}

\def\dt#1{{{\kern -.0mm\rm d}}#1\,}

\def\tr{{\rm Tr}}

\def\>{\rangle}
\def\<{\langle}

\def\ot{\otimes}

\begin{document}
\draft

\title{Thermodynamics of  quantum informational systems -
Hamiltonian description}

\author{Robert Alicki$^{1}$, Micha\l{} Horodecki$^{1}$,
Pawe\l{} Horodecki$^{2}$ and Ryszard Horodecki$^{1}$}

\address{$^1$ Institute of Theoretical Physics and Astrophysics,
University of Gda\'nsk, 80--952 Gda\'nsk, Poland,\\
$^2$Faculty of Applied Physics and Mathematics,
Technical University of Gda\'nsk, 80--952 Gda\'nsk, Poland
}

\begin{abstract}
It is often claimed, that from a quantum system of $d$ levels, and entropy $S$ and heat bath of temperature $T$
one can draw $kT(\ln d -S)$ amount of work. However,  the usual 
arguments basing on Szilard engine, are not fully rigorous. Here we prove the formula 
within Hamiltonian description of drawing work from a quantum system and a heat bath,
at the cost of entropy of the system. We base on the derivation of 
thermodynamical laws and quantities in [R. Alicki, J. Phys. A, {\bf 12}, L103 (1979)]
within weak coupling limit. Our result provides fully physical  scenario for
extracting thermodynamical work form quantum correlations
[Oppenheim et al. Phys. Rev. Lett. {\bf 89}, 180402 (2002)].
We also derive Landauer principle as a consequence of second law 
within the considered model.
\end{abstract}

\maketitle

\section{Introduction}
It is well known that thermodynamical  work can be drawn at 
the expense of entropy. An obvious example is a container with ideal gas placed within 
thermal reservoir at temperature T. Since the energy of ideal gas 
does not depend on volume, by expanding gas we draw work from reservoir solely at 
the expense of entropy. The drawn work is equal to $T \Delta S$. 
A more sophisticated example is Maxwell demon in Szilard engine \cite{Szilard}.
The engine consists of a box with  one particle of gas. 
The demon measures where the particle is, puts the piston and 
by expansion draws $kT \ln 2$ of work. The price is that 
entropy of demon increases by $1$ bit, as was pointed out 
by Bennett \cite{Bennett82} based on Landauer work \cite{Landauer}.  
The demon serves as entropy sink, which is needed to divide heat from 
the heat bath into entropy and work. Unlike the gas in container 
we deal here with microscopic objects: we have one molecule, and the demon is thought 
to be a small, microscopic being. Moreover, demon is not assumed to be necessarily
in equilibrium state, so that thermodynamical quantities are not applicable.   
This gives rise to the following general view:
if one has an $d$-level quantum system in a pure state $\psi$ 
one can draw $k T \ln d$ work out of heat bath of temperature $T$. If a state of the system 
is a mixed state $\rho$, then the amount of work would correspond to 
\be
W=k T (\ln d - S(\rho))
\label{work}
\ee
 where $S$ is von Neumann entropy. The latter function 
can be viewed as information contents of a state. 

Recently, the idea that one can draw work from  heat bath and a system with 
non-maximal entropy was used to investigate quantum properties 
of compound systems \cite{OHHH2001}. The equivalence between work 
and information (see \cite{Levitin93}) was  used in definition 
of the so called {\it work deficit}, which was a difference between
the work drawn when there is global access to the bipartite 
system, and when only local operations and classical communication 
is allowed \cite{note:zurek}. As a result, a new paradigm of investigation 
of correlations of quantum compound system was obtained,
in which quantum correlations manifest themselves through 
a {\it loss} of information during its concentration to local subsystems
for the purpose of drawing work from local heat baths. 
The paradigm, though based on thermodynamical ideas,
can be formulated  solely  by means of basic logical structure of 
thermodynamics, without the need of referring to the process of drawing 
work \cite{nlocc}. Yet, the connection with physical quantities such as 
energy, work and heat makes it even more interesting,
especially in the context of more and more realistic proposals to implement 
thermodynamical quantum microengines  
\cite{Lloyd-demon,Scully-negentropy,note:scully,Kosloff-microengines}.

However, so far in the literature 
there is no rigorous Hamiltonian description of a process of drawing work 
from heat bath and additional quantum system, to show 
that we can change information into work  according to formula (\ref{work}).
In this paper we provide such a description basing on derivation of phenomenological 
thermodynamics from theory of quantum open system \cite{Spohn78,Alicki79}.
We will then provide a physical description of protocols considered in \cite{OHHH2001}
for drawing work from local heat baths and compound systems by LOCC. 
We assume that the quantum system is coupled weakly to the bath. The assumption 
is needed to give rise to thermodynamical regime.
If the system is instead strongly coupled to the reservoir (see e.g. \cite{Nieuwenhuizen}), 
our results do not apply, then  however it is hard even to define work and heat, 
due to large fluctuations of energy of the system caused by interaction. 
In  the paper we also show how Landauer principle follows from second law derived 
within the model we discuss (c.f. \cite{Piechocinska}).

\section{Protocol of concentration of information to local form}
\begin{figure}
\vskip0.5cm
\includegraphics[scale=0.5]{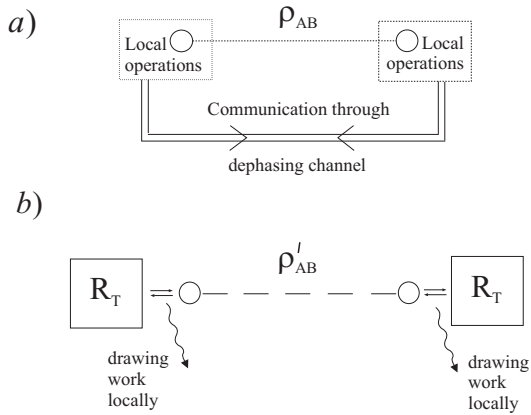}
\caption[system]{Protocol of drawing work from bipartite state 
and local heat baths.}
\label{fig:protocol}
\end{figure}

In this section we will describe the idea of \cite{OHHH2001} which motivates the present work.
There are  two parties, that are situated in distant labs. 
In each lab there is heat bath with temperature $T$. The parties  Alice and Bob,
possess subsystems $A$ and $B$ os a quantum system being in some state $\rho_{AB}$.
They can send particles to each other via fully decohering channel, 
i.e. the channel that removes off-diagonal terms  of a density matrix 
of the sent system in a fixed basis. Locally, their actions are not constrained. 
The task is to draw as much work as possible by use of the system they share 
and local heat baths. This requires to transform the 
initial state $\rho$ into such a final state $\rho'$, 
that its local entropies $S'_A$ and $S'_B$ are minimal.  
Such process can be called concentration of information to local form. 
Then Alice and Bob can apply local engines to draw amount 
$T[(S_A^{\max}-S_A')+(S_B^{\max}-S_B')$
of work (we have incorporated here Boltzmann constant to the entropy). In this way 
the question was as a matter of fact separated 
from such thermodynamical notions as work, heat, energy, 
and reduced to logical problem of 
minimal entropy production in the process of concentration of information. 
The described setup is illustrated on figure  \ref{fig:protocol}.

In those works, there was no detailed analysis of energy balance 
for the operations that transform the initial state into final state. 
Also there was no analysis of the final process of drawing work 
locally, that is drawing work from single heat bath and a system.

\section{Drawing work from heat bath and quantum system}
\subsection{Quantum system as a model of heat engine}

In this section we recall the microscopic model for heat engine of \cite{Spohn78,Alicki79}.
In particular, we will present thermodynamical quantities and laws within
dynamical model, where the working body is quantum system, 
that is driven by external force, and can be coupled to reservoirs. 
The laws are not postulated, but derived. It should be noted here, 
that the second law derived for quantum reservoir, but without a system as working body 
was derived in $C^*$-algebraic context by Pusz and Woronowicz \cite{Pusz-Woronowicz78}.

\begin{figure}[h]
\vskip0.5cm
\includegraphics[scale=0.5]{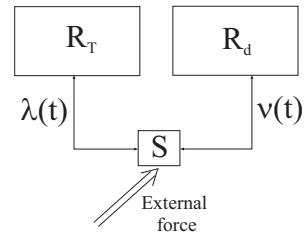}
\caption[system]{Quantum system driven by external force and interacting with reservoirs}
\label{fig:system}
\end{figure}

\paragraph{Quantum system and reservoirs.}
Consider a quantum system $S$ thermal  reservoir $R_T$, and decohering 
reservoir $R_d$. 
The state of the system is denoted by $\rho(t)$. The system  
is coupled to thermal bath via coupling constant $\lambda$ 
and to the decohering reservoir via constant $\nu$. The constants 
are external parameters, that can be changed, so that the system 
can be coupled to reservoirs at our will. The self Hamiltonian 
of the system can be changed in time by external force.  The whole setup 
is illustrated on figure \ref{fig:system}.

Thus the total Hamiltonian is of the form
\be
H_{S+R_T+R_d}= H_S(t)+ \lambda(t) H_{SR_T} + \nu(t) H_{SR_d}
\ee
We can single out three time scales: (1) $\tau_H$ -  characteristic scale of change of 
Hamiltonian $H_S$ and of coupling constants (2) $\tau_S\simeq \lambda^2,\nu^2$ - relaxation 
time of the system  and (3) $\tau_{R_d},\tau_{R_T}$ - relaxation times of 
reservoirs (for our purpose, both times,
as well as both coupling constants can be of the same order of magnitude). 
Assuming that $\tau_H,\tau_S\gg \tau_R$, the evolution of the system
can be approximated by Markovian master equation 
\cite{GorriniKossakowski76,Lindblad76,Davies76,Alicki-Lendi}.
\be
{\dt \rho\over \dt t}
=i^{-1}[H(t),\rho] + \lambda^2(t) L(t)\rho + \nu^2(t) K(t)\rho
\ee
where $L$ describes interaction with thermal reservoir while $K$ describes 
interaction with decoherence reservoir. The generators $L,K$ include the shifts 
of  Hamiltonians. They depend on time through change of coupling constants, 
that can switch them on and off and also through the change of Hamiltonian. 
Both generators $L$  and $K$
depend functionally on Hamiltonian: $L$ it causes relaxation of the system to 
the Gibbs state $\rho(t)=Z^{-1} e^{-\beta H(t)}$ while  the generator $K$ 
dephases state in basis of the self Hamiltonian $H(t)$. Thus the interaction with
thermal reservoir can change energy of the system while the decohering 
reservoir does not change the energy, only destroys coherences between
the eigenstates of self-Hamiltonian. Once the Hamiltonian is 
switched off, $K$ decoheres in some  basis determined 
by interaction Hamiltonian. 

Usually, it is not the case that there are two separate 
reservoirs. A typical reservoir which can be described 
by Markovian master equation is rare gas, and it causes both effects: The suitable generator 
can be divided into two parts, decohering one $K$ and  $L$ inducing transitions 
between the levels. When the Hamiltonian is switched on, both parts are  present,
while for degenerated levels, only the decohering part is present. 
In such case, one can control couplings as follows:
in order to switch off the $L$ part, one has to switch off the Hamiltonian.
To effectively switch off both parts, one should simply use much faster changes 
of Hamiltonian than the time of decay induced by reservoir. 
It is not possible to have $L$ but not $K$, however it is not important in 
the present context.

\paragraph{Thermodynamical quantities and laws.}
Now one can define the thermodynamical quantities as follows. The thermodynamical 
energy of the system is identified with average energy 
of the system 
\be
E(t)=\tr \rho(t) H(t)
\ee
Heat and work are defined as follows
\ben
&&Q(t)=\int_{0}^{t} \tr \left[ {\dt \rho(t) \over \dt t} H(t) \right] \dt t\\ \nonumber
&& W(t)=\int_{0}^{t} \tr \left[\rho(t) {\dt H(t)\over \dt t}\right] \dt t. 
\een
They obviously satisfy the first law of thermodynamics
\be
\dt E=\dt W +\dt Q
\ee
It is convenient to require that  the energy 
of the system does not change in time and thus  can  be set to be zero
\be
\tr \rho(t) H(t)= 0,
\ee 
so that $\dt W= -\dt Q$ in our case. The entropy is given by 
\be
S(\rho)=-k\tr \rho\ln \rho
\ee
The variation of entropy  can be divided into the part due to heat exchange, and 
the rest, called entropy production. 
\be
{\dt S \over \dt t }=  {1\over kT} {\dt Q \over \dt t} + \sigma(t)
\ee
The first part can be negative or positive, 
while the second one, defined by the equation, is always nonnegative
\be
\sigma(t)\geq 0
\label{eq:second-law}
\ee
as follows from monotonicity of relative entropy under physical processes 
\cite{Spohn78,McAdory77}.  
This is actually a statement of the Second Law.  
From the above formulas one gets  formula for work in any process 
\be
W=kT [(S(t_2) - S(t_1) - \int_{t_1}^{t_2}\!\!\!\sigma(t)\, \dt t]. 
\label{eq:optimal-work}
\ee

\subsection{Elementary processes} 

To show that from pure qubit one can draw work equal to $kT \ln 2$,
we have to show process that will change entropy from zero to maximal
equal to $\ln2$ with zero entropy production. To this end we will examine 
elementary processes which can be run. 

\begin{figure}
\includegraphics[scale=0.7]{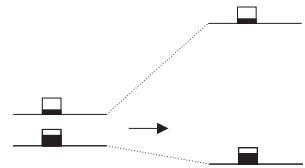}
\vskip.9cm
\caption{Adiabatic change of Hamiltonian. Black squares denote population,
which is unchanged during the process.}
\label{fig:adiab}
\end{figure}

\noindent
{\bf Adiabatic change of Hamiltonian.} In this process, the system is not coupled 
to the thermal bath, so that $\lambda(t)=0$. Also it is not coupled to 
the decohering bath i.e. $\nu(t)=0$. Since the operation is unitary, the entropy 
production $\sigma(t)$ is zero.  As mentioned, we allow only such changes 
that keep the energy to be zero. Thus, if the system initially 
had two degenerated levels, and it is in the state 
with populations $p_1$ and $p_2$ then we change Hamiltonian in such 
a way, that one of the level gets energy $+E/{2p_1}$ while the other $-E/{2p_2}$,
so that the change produces energy difference $E$, 
but the total energy is still zero. The Hamiltonian commutes with $\rho$ 
all the time: $[H(t),\rho]=0$.  The process is shown on figure \ref{fig:adiab}.

\begin{figure}
\vskip0,5cm
\includegraphics[scale=0.7]{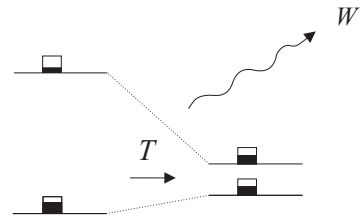}
\vskip0,5cm
\caption{Isothermal contact with reservoir. The population is changed: the final state 
has greater entropy than the initial one.}
\label{fig:isothermal}
\end{figure}

\noindent
{\bf Isothermal and quasistatic contact with thermal bath.} 
During the process, the state of the system is in equilibrium 
with thermal bath
\be
\rho(t)=Z^{-1} e^{-{H(t)\over kT}}
\ee
In this process the system is coupled to the reservoir, so that $\lambda(t)\not = 0$.
Still $\nu=0$, as we do not couple the system to decohering reservoir.
The process is quasistatic, which means that the time $\tau_H$ 
is mush longer than the time of system relaxation $\tau_S$. This means that 
the changes of Hamiltonian are so slow, that the system is all the time 
approximately in equilibrium state. Therefore
The entropy production is all the time zero while entropy production 
can be only due to either decohering reservoir or nonequilibrium processes. 
The process is illustrated on figure \ref{fig:isothermal}.

\begin{figure}
\includegraphics[scale=0.5]{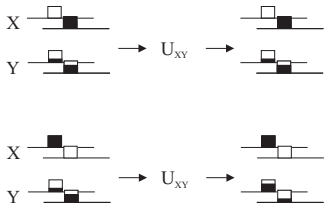}
\vskip0,5cm
\caption{Unitary c-not gate. When the qubit X is in ground state 
Y is unchanged, otherwise Y is flipped.}
\label{fig:cnot}
\end{figure}

\noindent
{\bf Unitary gates.} The system is not coupled to any reservoir ($\lambda,\nu=0$),
and the initial and final Hamiltonian is equal to $0$:
\be
H(t_1)=H(t_2)= 0
\ee
Thus, the Hamiltonian is switched on to run required unitary operation,
and then switched off. Thus the unitary gate is actually composed of two adiabatic 
changes of Hamiltonian. As it should be, the work  performed during unitary 
gate is zero, as it is equal to heat exchange, so that 
\ben
&&W=\int_{t_1}^{t_2} \tr \left\{ {\dt \rho(t) \over \dt t} H(t) \right\} \dt t=\\
&&\int_{t_1}^{t_2} \tr \left\{-i \hbar [\rho(t),H(t)] H(t) \right\} \dt t =0 \nonumber
\een
The most common example of two-qubit unitary gate is c-not gate,which is defined by 
\ben
U_{XY} |0\>_X|0\>_Y =|0\>_X|0\>_Y \\ \nonumber
U_{XY} |0\>_X|1\>_Y =|0\>_X|1\>_Y \\ \nonumber
U_{XY} |1\>_X|0\>_Y =|1\>_X|1\>_Y \\ \nonumber
U_{XY} |1\>_X|1\>_Y =|1\>_X|0\>_Y \\ 
\een
The transformation if applied to standard basis 
does not change first qubit $X$. Second qubit $Y$ is also untouched once 
the first qubit is in state $|0\>$ and is flipped if the first qubit is in state 
$|1\>$ (see Figure \ref{fig:cnot}). 
The qubit $X$ is called {\it source}, while the qubit 
$Y$ - {\it target}. The gate can be realized by applying 
the following Hamiltonian 
\be
H(t)=E(t) |1\>_X\<1| \otimes |-\>_Y\<-|,  
\label{eq:hamil-cnot}
\ee
with $\int E(t) \dt t=\pi$, $|-\>={1\over \sqrt 2} (|0\> +|1\>)$.

{\bf Irreversible pure decoherence.}
The system is coupled  to the decohering reservoir only, so that $\lambda=0$ 
but $\nu\not=0$. The Hamiltonian is off i.e. $H(t)=0$. Under such process, the 
off diagonal terms in a fixed basis disappear. Of course, to draw optimal 
work one will not use this process. It is useful to model communication
via dephasing channel: Instead of sending qubit via the channel,
Alice can first decohere it locally, and then send such decohered intact to Bob.

\subsection{Drawing work by use of quantum system}
\label{subsec:work-pure}
\begin{figure}[htb]
\vskip0,5cm
\includegraphics[scale=0.8]{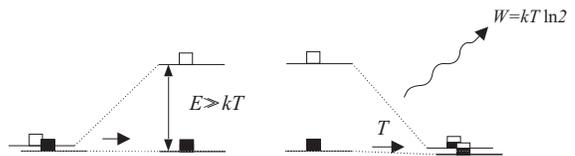}
\vskip0,5cm
\caption{Drawing work by use of a pure qubit. Dotted line denotes energy 0 level.
The initial state is pure, occupating ground level, while the final state 
has both levels equally populated.}
\label{fig:bitwork}
\end{figure}

Consider the following initial conditions: Hamiltonian is equal to zero, and the 
state is pure.  As we know, to draw work without entropy production, 
one should apply isothermal process, which requires the system 
to be in equilibrium state. However the state is pure, and Gibbs state is always mixed. 
Nevertheless  one can obtain almost Gibbs state.
To this end one performs adiabatic switching on the Hamiltonian in such a way,
that the state is ground state of the Hamiltonian and  the obtained energy splitting 
must be much higher than $kT$. Then to a good approximation, the pure state 
is equal to Gibbs state. Subsequently, one switch on isothermal contact with 
reservoir, with the Hamiltonian adiabatically changed to zero. 
Therefore, since all the time state is in equilibrium, the final state is equally populated.
The obtained work is then equal to $kT \ln 2$ (the initial entropy was 0, 
while the final one is $\ln 2$).  The process is illustrated on figure \ref{fig:bitwork}.
If the initial qubit is not in pure state but in some mixed state $\rho$ 
the amount of work is given by its  information (or negentropy) contents 
\be
W=T(S_{\max}-S(\rho))=kT \ln 2 - TS(\rho).
\ee

\section{Deriving Landauer principle}
From the formula for work (\ref{eq:optimal-work}) implied directly by the second law
(\ref{eq:second-law}), we can derive Landauer principle. 
Recall, that the latter says that erasure  one bit of information in contact with heat 
reservoir 
with temperature $T$ costs  dissipation of $kT\ln$ of energy. 
The information is here considered in Shannon {\it subjective} sense, where 
information produced by a source 
is measured by the entropy of ensemble emitted by the source. Thus 1 bit of information 
is represented by a 1-bit register in maximally mixed state. Quantum mechanically, it is 
one qubit in maximally mixed state. The "erasure of the bit" means transformation 
of the state 
into some standard pure state $|0\>$. Such state represents zero information in Shannon 
sense, because it is apriori known. In terms of {\it objective} information understood 
as purity 
or negentropy, we have the converse interpretation: the initial state, maximally mixed,
hence with zero information contents, become transformed into pure state,
containing 1 bit o information. Then we could say: creation of information 
costs dissipation of $kT \ln 2$ of energy. (of course the information 
of the total system: qubit plus reservoir is conserved, as the entropy of the 
reservoir increased). 

Let us now prove the Landauer principle within Hamiltonian model.
Assume that one has a maximally mixed qubit. Denote by $W_{er}$ 
the work that needs to be dissipated into reservoir, 
while bringing the system to the pure state $|0\>$. 
If less than $k T \ln2$ work would be dissipated, then we could draw work 
from system in maximally mixed state and heat bath, 
what is forbidden by formula (\ref{eq:optimal-work}). Indeed,
one could start with maximally mixed qubit, 
then erase it to pure state by $W_{er}< kT \ln 2$ 
and then by applying isothermal quasistatic process, draw $W=kT \ln 2$.
The final state of qubit is again maximally mixed, but
one would have excess of work $W-W_{er}>0$. 
Thus to erase a single bit, one needs to dissipate at least $kT\ln 2 $ 
of work. To see that it is optimal, it is enough to consider reverse 
of process drawing work from previous section.
 
\section{Drawing work by local engines in distant lab paradigm}
In this section we will treat the process of Ref. \cite{OHHH2001} 
depicted on figure \ref{fig:protocol}. It consists of two stages:
a) concentration of information to local form, b) local drawing work. In previous section 
we have provided Hamiltonian description of the second process. 
Thus form a state $\rho'_{AB}$ and local reservoirs Alice and Bob are able
to draw amount of work $W=T [(S_A^{\max}-S(\rho'_A))+(S_B^{\max}-S(\rho')_B)]$. 
Thus it is enough to 
consider the stage a) and to show that energy consumption of this stage 
is negligible. The protocol consist of (i) local unitary gates, (ii) local 
contact with decohering reservoir and (iii) sending quantum systems 
between the labs. As argued in previous section, the unitary gates 
can be done without energy consumption. The interaction with decohering reservoir 
does not change energy. The last step is actually an operation of 
transport, which in conservative potential does not uses up energy. 

Let us present an example of the stage of localisation of information from 
two two-level quantum systems in maximally  entangled state   of the form:
\be
\psi_+={1\over \sqrt2}(|0\>_A 0\>_B+|1\>_A |1\>_B)
\label{singlet}
\ee
(c.f. \cite{OHHH2001}).
We assume that the levels are degenerated  unless by external field we switch 
on some Hamiltonian (example is  spin with driving external force being magnetic field). 
The state has maximally mixed subsystems, so that Alice and Bob 
cannot draw  any work locally. Thus in first 
stage, Alice and Bob will aim to concentrate information to local subsystems.
To this end, they take a third two-level quantum system  $C$ which will be used 
for communication. The particle is prepared in pure state $|0\>$ 
and after the whole protocol it must be return in such state 
(thus it serves for a working body, for which the cycle has  
to be closed). Before and after the process, the system $C$  is in equilibrium 
with the bath, and its pure state is maintained by using potential with gap between 
ground and first excited state much greater than $kT$ as explained in 
sec. \ref{subsec:work-pure}.
To start the process of localisation of information, Alice and bob will switch 
off the potential,
and the process the system $C$ is not coupled to the thermal reservoir. After the process, 
the qubit $C$ can be again put into high potential well, and put in contact with reservoir.

The process of localisation of information of the state (\ref{singlet})
is the following. 
\def\cbit{C}
\bee
\item The particle $\cbit$ is coupled to the particle $A$ via 
Hamiltonian  of equation (\ref{eq:hamil-cnot})
with source system $X=A$ and target $Y=\cbit$.
This realizes c-not gate between $A$ and $\cbit$ 
which results in transition 
\ben
&&|0\>_{\cbit}{1\over \sqrt2}(|0\>_A 0\>_B+|1\>_A |1\>_B) \to \\ \nonumber
&&\to {1\over \sqrt2}(|0\>_{\cbit}|0\>_A  |0\>_B+|1\>_{\cbit}|1\>_A  |1\>_B)
\een
The qubit $\cbit$ (target) has measured the qubit $A$ (source).
\item In LOCC paradigm, $\cbit$ is classical bit, therefore Alice 
will switch on the decohering reservoir for the qubit $\cbit$. The state 
turns then into probabilistic mixture:
\ben
&&{1\over \sqrt2}(|0\>_{\cbit}|0\>_A  |0\>_B+|1\>_{\cbit}|1\>_A  |1\>_B )\to \\ \nonumber
&&\to{1\over 2} |0\>_{\cbit}\<0| \ot|0\>_A\<0|\ot|0\>_B\<0|+\\ \nonumber
&&+{1\over 2}
|1\>_{\cbit}\<1|\ot |1\>_A\<1|\ot |1\>_B\<1|
\een
\item The qubit ${\cbit}$ is communicated to Bob. The state is now
\ben
&&{1\over 2} |0\>_A\<0|\ot|0\>_B\<0|\ot|0\>_{\cbit}\<0| +\\ \nonumber
&&+{1\over 2}
 |1\>_A\<1|\ot |1\>_B\<1| \ot|1\>_{\cbit}\<1|
\een
\item Bob applies Hamiltonian of (ref{eq:hamil-cnot}) 
with $\cbit$ being source and 
 $B$  - target system. As a result the qubit $B$ 
uncouples from two other ones, and the total state is 
\ben
&&{1\over 2} \bigl(|0\>_A\<0|\ot |0\>_{\cbit}\<0| +
|1\>_A\<1|\ot |1\>_{\cbit}\<1|\bigr) \ot  \\ \nonumber
&& 
\ot |0\>_B\<0|
\een
\item The qubit $\cbit$ is sent back to Alice. 
Alice applies  the same Hamiltonian as at the beginning 
($H_{XY}$ of eq. (\ref{eq:hamil-cnot}) with $X=A$, $Y= \cbit$ ) to finish cycle
by resetting the qubit $\cbit$ to the standard state $|0\>$. The final state is:
\be
|0\>_{\cbit}\<0|\ot {1\over 2} \bigl(|0\>_A\<0| + |1\>_A\<1|\bigr)
\ot |0\>_{B}\<0|   \\ \nonumber
\ee
\eee
This was the stage of localization of information. 
During the process, the entropy increased by one bit. 
The Bob's qubit is now pure, and $kT\ln 2$ work can be drawn from it 
by adiabatic quasistatic coupling to thermal reservoir. 
In this process next bit of entropy is produced. 
The initial pure state $\psi_{AB}$ becomes maximally mixed. 

\section{Conclusions}
In conclusion, we have  presented a Hamiltonian model of drawing work from single heat 
bath and a quantum system. 
In the model the system is weakly coupled to the heat bath. In the process whole 
heat is changed into work; the second law is saved, 
because the quantum system increases its entropy. The overall process can be viewed 
as depleting entropy from the noisy energy 
contained in  Gibbs state. The latter state contain both entropy. In the process of 
drawing work the entropy 
goes to qubit, while the energy is obtained in ordered form (it can be stored 
as potential energy.

We have pointed out how Landauer principle follows from the second law derived 
within the model: were it possible 
to erase qubit at lower cost than $kT \ln 2$, one could draw more work from pure qubit, 
which would violate second law.

Finally we have represented in Hamiltonian picture the process of drawing work 
from compound systems  and local heat baths
by local operations and classical communication of \cite{OHHH2001}. Such process consists 
of two stages: localisation of information (which is usually irreversible) and 
then drawing work 
locally. We have carried out energy balance of such process, and 
obtained that the process of localisation of information no energy needs to be spent. 
In the second stage, by use of our model, one can draw the amount of fork determined 
by local information contents. 
Thus the localisable information of \cite{OHHH2001,nlocc} can be interpreted as 
the maximal amount of work 
drawn by use of quantum system distributed into distant labs and  local heat 
baths  operated  by local 
operations and classical communication.

The authors would like to thank  Chris Fuchs, Jonathan Oppenheim and  
Jos Uffink for interesting discussions. 
This paper has been supported by the Polish Ministry of Scientific
Research and Information Technology under the (solicited) grant No.
PBZ-MIN-008/P03/2003 and by EC grants RESQ, Contract No. IST-2001-37559 
and  QUPRODIS, Contract No. IST-2001-38877.

\bibliography{refmich,silnik}

\begin{thebibliography}{22}
\bibitem[{not({\natexlab{a}})}]{note:zurek}
\bibinfo{note}{The difference between drawing work globally and locally was
  subsequently investigated by \.Zurek \cite{Zurek-demons-02}.}
\bibitem[{not({\natexlab{b}})}]{note:scully}
\bibinfo{note}{It is not quite clear whether in \cite{Scully-negentropy} the
  energy balance is properly carried out.}
\end{thebibliography}
\bibliographystyle{apsrev}

\end{document}